\documentclass[twocolumn]{aastex631}

\usepackage{amsmath}
\usepackage{hyperref}

\newcommand{\pr}{\ensuremath{P_{\rm r}}}

\newcommand{\pq}{\ensuremath{P_Q}}
\newcommand{\pu}{\ensuremath{P_U}}

\newcommand{\ainv}{\ensuremath{\alpha_{0}}}
\newcommand{\amin}{\ensuremath{\alpha_{\rm{min}}}}

\newcommand{\pmin}{\ensuremath{P_{\rm{min}}}}

\begin{document}

\title{Multi-Wavelength Post-Perihelion Polarimetry of Interstellar Comet 3I/ATLAS}

\author[0000-0002-6610-1897]{Zuri Gray}
\affiliation{Department of Physics, University of Helsinki, PO Box 64, 00014, Finland}

\author[0000-0002-4516-459X]{Galin Borisov}
\affiliation{Institute of Astronomy and National Astronomical Observatory, Bulgarian Academy of Sciences, 72 Tsarigradsko Chauss\'ee Blvd., BG-1784 Sofia, Bulgaria}
\affiliation{Armagh Observatory \& Planetarium, College Hill, Armagh, BT61 9DG, UK}

\author[0000-0002-9321-3202]{Ludmilla Kolokolova}
\affiliation{Department of Astronomy, University of Maryland, College Park, MD 20742-2421, US}

\author[0000-0001-5989-3630]{Johannes Markkanen}
\affiliation{Institut fur Geophysik und Extraterrestrische Physik, TU Braunschweig, Mendelssohnstr. 3, D-38106 Braunschweig, Germany}

\author[0000-0002-8122-3606]{Yuna G. Kwon}
\affiliation{Armagh Observatory \& Planetarium, College Hill, Armagh, BT61 9DG, UK}

\author[0000-0002-7156-8029]{Stefano Bagnulo}
\affiliation{Armagh Observatory \& Planetarium, College Hill, Armagh, BT61 9DG, UK}

\author[0000-0002-6645-334X]{Alberto Cellino}
\affiliation{INAF -- Osservatorio Astrofisico di Torino, I-10025 Pino Torinese, Italy}

\author[0000-0002-8910-1021]{Rosemary C. Dorsey}
\affiliation{Department of Physics, University of Helsinki, PO Box 64, 00014, Finland}

\author[0000-0002-8418-4809]{Grigori Fedorets}
\affiliation{Finnish Centre for Astronomy with ESO, University of Turku, FI-20014 Turku, Finland}
\affiliation{Department of Physics, University of Helsinki, PO Box 64, 00014, Finland}

\author[0000-0002-5624-1888]{Mikael Granvik}
\affiliation{Department of Physics, University of Helsinki, PO Box 64, 00014, Finland}
\affiliation{Asteroid Engineering Laboratory, Lule\r{a} University of Technology, Box 848, SE-98128 Kiruna, Sweden}

\author[0000-0002-9870-123X]{Eric MacLennan}
\affiliation{Department of Physics, University of Helsinki, PO Box 64, 00014, Finland}

\author[0000-0001-7403-1721]{Antti Penttil\"a}
\affiliation{Department of Physics, University of Helsinki, PO Box 64, 00014, Finland}

\author[0000-0001-8058-2642]{Karri Muinonen}
\affiliation{Department of Physics, University of Helsinki, PO Box 64, 00014, Finland}

\author[0000-0002-5138-3932]{Olga Mu\~{n}oz}
\affiliation{Instituto de Astrof\'{\i}sica de Andaluc\'{\i}a, CSIC, Glorieta de la Astronomia s/n, E-18008 Granada, Spain\\}

\author[0000-0002-4278-1437]{Philippe Bendjoya}
\affiliation{Université Côte d’Azur, Laboratoire Lagrange, OCA, CNRS UMR 7293, Nice, France}

\author[0000-0002-6509-6360]{Maxime Devogèle}
\affiliation{ESA NEO Coordination Centre, Largo Galileo Galilei 1, I-00044 Frascati (RM), Italy}

\author[0000-0001-8694-9038]{Simone Ieva}
\affiliation{INAF -- Osservatorio Astronomico di Roma, 00078 Monte Porzio Catone (RM), Italy}

\author[0000-0002-0246-4648]{Francisco J. García-Izquierdo}
\affiliation{Instituto de Astrof\'{\i}sica de Andaluc\'{\i}a, CSIC, Glorieta de la Astronomia s/n, E-18008 Granada, Spain\\}


\begin{abstract}
We present new post-perihelion polarimetric observations of the third discovered interstellar object, 3I/ATLAS (C/2025 N1; hereafter 3I), obtained using FORS2 at the Very Large Telescope, ALFOSC at the Nordic Optical Telescope, and FoReRo2 at the 2 m Ritchey-Chrétien–Coudé telescope. The observations span phase angles 1--30$^{\circ}$, providing the most extensive polarimetric phase angle coverage obtained for an interstellar object to date. The post-perihelion measurements reveal that the unusual polarimetric properties of 3I persist across perihelion and match recent independent post-perihelion observations, indicating no significant evolution in the polarimetric properties. The increased phase angle coverage allowed us to further constrain the minimum polarisation to $-2.9\%$ at phase angle $5.5^{\circ}$. Multi-band BVRI observations reveal a wavelength dependence of polarisation, with the negative polarisation branch becoming deeper and shifted toward smaller phase angles at longer wavelengths. The polarimetric colour is predominantly red and increases with phase angle, consistent with behaviour observed in other comets. Imaging and polarimetric maps show a substantially more extended coma after perihelion, while the polarisation distribution itself remains smooth and spatially homogeneous. Numerical modelling suggests that the unusually deep polarisation phase curve originates from moderately porous dust aggregates consisting of weakly absorbing sub-micron monomers, resulting in much brighter dust than that in typical solar system comets. 

\end{abstract}

\section{Introduction}
\label{Intro}

3I/ATLAS (C/2025 N1; hereafter `3I'), discovered on 2025 July 1 by the ATLAS sky survey, represents the third ever discovered interstellar object (ISO), preceded by 1I/'Oumuamua in 2017 and 2I/Borisov in 2019. At discovery, 3I was inbound towards the inner Solar System at a heliocentric distance of $r_{\rm{H}} = 4.5$ au, providing the astronomical community with the exceptional opportunity to probe the composition, properties and evolution of interstellar material across an entire perihelion passage. Consequently, 3I quickly became the target of extensive observational campaigns across a wide range of wavelengths and techniques, including photometry, spectroscopy, and polarimetry.  

Pre-perihelion polarimetric observations by \citet{gray2025ApJ} revealed that 3I exhibits unusual polarimetric properties. The polarimetric phase curves of small solar system bodies, comets and asteroids, are characterised by a negative polarisation branch (NPB) at small phase angles and a positive polarisation branch (PPB) at larger phase angles, where the polarisation vector is oriented parallel and perpendicular to the scattering plane, respectively. Historically, solar system comets (hereafter `SSC') generally fit into one of two polarimetric classes: high- or low-polarisation comets \citep{kiselev2015}. Only a small number of comets deviate significantly from these trends, most notably C/Hale-Bopp (C/1995 O1) \citep{hadamcik2003} and 2I/Borisov \citep{bagnulo2021Natur}, which are characterised by polarisation values systematically higher than high-polarisation comets and are mutually consistent within their overlapping phase angle range.

The pre-perihelion polarimetric campaign of 3I, spanning phase angles of $\alpha = 7.7-22.4^{\circ}$, sampled both the NPB and PPB. The resulting polarimetric phase curve revealed a deep and narrow NPB, reaching a minimum polarisation of $\pmin = -2.7\%$ at $\amin=6.8^{\circ}$, and an inversion angle (phase angle at which polarisation switches sign) of $\ainv = 17.1^{\circ}$. These values are notably different from those of typical SSCs, characterised by $\pmin \sim -1.5\%$ at $\amin \sim 10^{\circ}$, and $\ainv\sim 20-22^{\circ}$ \citep{kiselev2015}. This polarimetric behaviour distinguishes 3I not only from the bulk of SSCs, but also from the unique comet-polarisation class populated solely by C/Hale-Bopp (C/1995 O1) and 2I/Borisov, the latter being only one other ISO studied polarimetrically. These pre-perihelion measurements suggest the presence of large particles composed of a mixture of dark and bright (or icy) material \citep{gray2025ApJ,ivanova2023}.


While the pre-perihelion results already established 3I as a polarimetrically unusual object, they also raised an important question: do the peculiar scattering properties of 3I evolve as the comet progresses through perihelion passage and experiences a variety of environmental conditions? Indeed, numerous studies have reported substantial evolution in the comet's activity and volatile composition across perihelion (e.g. \citealt{cordiner2025,combi2026,tan2026,zhao2026,medler2026,lisse2026}), motivating the search for corresponding changes in 3I's polarimetric properties.

In this paper, we present new post-perihelion multi-wavelength polarimetric observations of 3I, extending the phase angle coverage to small, previously uncovered phase angles. To date, two independent groups have published post-perihelion polarimetric observations of 3I, reaching different conclusions regarding the evolution of its dust properties \citep{zheltobryukhov2025,choi2026}. We compare our measurements to the pre-perihelion polarisation of the comet \citep{gray2025ApJ}, as well as the other post-perihelion observations, and explore the implications of our results through dedicated light scattering modelling.

\section{Observations and Analysis}\label{Sect_Observations}
\subsection{Polarimetry}
We obtained post-perihelion polarimetric observations of 3I using the same three facilities as \citet{gray2025ApJ}: FORS2 at the Very Large Telescope (VLT; 8.2 m) \citep{appenzeller1998}, FoReRo2 at the 2m Ritchey-Chrétien-Coude (RCC) \citep{FoReRo2OLD,FoReRo2NEW}, and ALFOSC at the Nordic Optical Telescope (NOT; 2.56 m). The observing campaign consisted of 18 epochs 30--101 days after perihelion (2025 Oct 29), corresponding to phase angles $\alpha$ = 30.4--1.0$^\circ$ and heliocentric distances $r_H$ = 1.7--3.9 au. Our observing log is summarised in Table \ref{tableLog}.

We obtained multi-band BVRI observations with FORS2 in all but two epochs, when only R-band data were taken. Additional R-band observations were carried out with FoReRo2 and ALFOSC. The transmission curves of all filters used in this campaign are given in the Appendix. For the FORS2 B- and I-band observations, we used narrowband filters in all but the final two epochs, where we adopted broadband filters to maintain multi-band coverage as the comet became fainter and the remaining telescope time became limited. Although such difference in filter bandpass may be important in cases where the polarisation varies strongly with wavelength, the polarisation of 3I changes smoothly across the optical range. As a result, any effects introduced by the differing filter transmission curves are expected to be negligible compared to photon noise uncertainties. Similarly, we expect differences in the bandpass of R-band filters used by FORS2, FoReRo2, and ALFOSC to have negligible wavelength-related effects on the polarisation measurement. 

The observing strategy and data reduction procedures follow those described in \citet{gray2025ApJ}, and thus, we refer the reader to this article for such details. The one difference between the pre- and post-perihelion campaigns is the ALFOSC polarimetric configuration: the latter ALFOSC observations were taken using the Wedged Double Wollaston (WeDoWo) configuration for single-shot polarimetry as opposed to using the FAPOL polarimetric setup. The specifics of the WeDoWo setup and observations are detailed in the Appendix. 

For the analysis of our data, we measure the degree of linear polarisation within a circular aperture of radius $\sim$~2,000--6,000~km centred on the photocentre. In addition, we construct polarimetric and imaging maps from the FORS2 data to investigate the coma morphology and the spatial distribution of polarisation. We note that the observations on 2026 January 4 were acquired at a Moon--target angular separation below the usual $30^{\circ}$ threshold. Therefore, we treat these measurements with somewhat greater caution, as increased scattered moonlight may contribute additional background and systematic uncertainty. 

\subsection{Gas contamination}
Light coming from cometary atmospheres is a superposition of radiation scattered by dust particles in the coma and that emitted by cometary gases. Since molecular fluorescence generally exhibits a lower polarisation degree compared to that of light scattered by cometary dust, the presence of gaseous emissions within the bandwidth of a given filter acts to dilute the measured degree of polarisation \citep[e.g.,][]{kiselev2004, jockers2005}. Consequently, the observed value of polarisation may be underestimated compared to the intrinsic polarisation of the dust coma. 

Since no simultaneous spectroscopic observations were obtained together with our polarimetric measurements, we assess the possible level of gas contamination within the filters used in our study using contemporaneous spectroscopic observations reported in the literature. As an independent consistency check, we also applied the approximate image-subtraction technique described by \citet{ishiguro2016}, and subsequently applied by \citet{kwon2017,kwon2018}, to estimate the gas contribution from our broadband images alone. The details of this technique are outlined in the Appendix. 

Numerous post-perihelion spectroscopic studies have shown that 3I exhibits molecular emission bands commonly observed in cometary spectra, including CN, $\rm{C_3}$, $\rm{C_2}$, together with atomic emission lines \citep{kawakita2026,zhao2026,medler2026,hoogendam2026}. The strongest emission feature, CN near 388~nm, as well as the $\rm{C_3}$ band near 405~nm, lay safely outside the wavelength coverage of the filters used in this work. The principal source of possible contamination in our data is the $\rm{C_2}$ Swan system between approximately 460--520~nm, whose strongest bands partially overlap the transmission wings of the B- and V-filters used here. Additional contamination in typical comet spectra can occur in the R-band due to weak $\rm{NH_2}$. However, 3I has been reported to be strongly NH$_2$ depleted \citep[e.g.,][]{kawakita2026}, suggesting that contamination in the R-band should be negligible. 

Multi-epoch spectroscopy by \citet{zhao2026} show that the molecular emission bands weaken steadily relative to the continuum with increasing heliocentric distance after perihelion. Using the transmission curves of the filters employed in this work, \citet[private communication]{zhao2026} estimate that gaseous emission contributes only 2--3\% of the total flux in the B- and V-bands on 2025 December 17 and 19, and drops to 0--1\% in both bands by 2026 January 11 and 20. These estimates indicate that gas contamination in our observed wavelength range is already small by mid-December, and becomes negligible by January. Regarding aperture sizes, the 2.3--2.5$''$ slit width used in \citet{zhao2026} is comparable to the aperture sizes used for our polarimetric measurements. 

Applying the image-based technique \citep{ishiguro2016}, we estimate that gas emission contributes of order 5--10\% in the B- and V-filters on December 18, and $<5\%$ on December 26. For the later epochs, the method did not yield robust constraints on the gas contribution, most likely because the gas emission had become too weak to be distinguished reliably from the dust continuum. These estimates are somewhat larger than those derived by \citet{zhao2026}. The image-based approach relies on an approximate separation of dust and gas using broadband images, whereas the spectroscopic estimates are obtained by directly integrating continuum-subtracted spectra over the measured filter transmission curves. Therefore, we consider the spectroscopic estimates as the more reliable determination of the gas contribution, while our image-based results should be interpreted as order-of-magnitude estimates. Nevertheless, both approaches consistently indicate that gas contamination was modest during our earliest observing epochs and became negligible as the comet receded from the Sun. 

Overall, measurements indicate that gas contamination within our observed wavelength range is small or negligible throughout our observing campaign. Thus, we regard the measured polarisation across all BVRI bands as representative of 3I's dust coma. 


\section{Results}\label{Sect_Results}

\begin{table*}
\caption{\label{tablePQ} Results of aperture polarimetric measurements of 3I using aperture radius between approximately 2000--6000 km.}
\centering
\footnotesize
\begin{tabular}{ccccr@{$\pm$}lr@{$\pm$}lr@{$\pm$}lr@{$\pm$}lr@{$\pm$}lr@{$\pm$}lr@{$\pm$}l}
\hline\hline
Date & Instrument & $\alpha$ & $r_H$ & \multicolumn{8}{c}{\pq\ (\%)} & \multicolumn{6}{c}{PC  (\%/100~nm)} \\
     &            & (deg)    & (au) 
     & \multicolumn{2}{c}{B} 
     & \multicolumn{2}{c}{V} 
     & \multicolumn{2}{c}{R} 
     & \multicolumn{2}{c}{I} 
     & \multicolumn{2}{c}{PC(B,R)}
     & \multicolumn{2}{c}{PC(V,R)}
     & \multicolumn{2}{c}{PC(R,I)} \\
\hline

2025-Nov-28.27 & ALFOSC & 30.4 & 1.7 
& \multicolumn{2}{c}{---}
& \multicolumn{2}{c}{---}
& 6.41 & 0.87
& \multicolumn{2}{c}{---} 
& \multicolumn{2}{c}{---} & \multicolumn{2}{c}{---} & \multicolumn{2}{c}{---} \\

2025-Dec-17.34 & FORS2 & 25.5 & 2.2
& \multicolumn{2}{c}{---}
& 3.96 & 0.23
& 4.32 & 0.25
& 4.90 & 0.22 
& \multicolumn{2}{c}{---}
& 0.36 & 0.35 
& 0.32 & 0.19 \\

2025-Dec-18.34 & FORS2 & 24.9 & 2.3
& 3.07 & 0.32
& 3.45 & 0.22
& 3.93 & 0.17
& 4.73 & 0.21 
& 0.51 & 0.23 & 0.49 & 0.29 & 0.45 & 0.15 \\

2025-Dec-19.10 & ALFOSC & 24.4 & 2.3
& \multicolumn{2}{c}{---}
& \multicolumn{2}{c}{---}
& 3.17 & 0.34
& \multicolumn{2}{c}{---} 
& \multicolumn{2}{c}{---} & \multicolumn{2}{c}{---} & \multicolumn{2}{c}{---} \\

2025-Dec-26.08 & ALFOSC & 19.5 & 2.5
& \multicolumn{2}{c}{---}
& \multicolumn{2}{c}{---}
& 1.36 & 0.26
& \multicolumn{2}{c}{---} 
& \multicolumn{2}{c}{---} & \multicolumn{2}{c}{---} & \multicolumn{2}{c}{---} \\

2025-Dec-26.32 & FORS2 & 19.3 & 2.5
& 0.43 & 0.09
& 0.73 & 0.06
& 1.22 & 0.05
& 1.74 & 0.06 
& 0.47 & 0.06 & 0.51 & 0.08 & 0.28 & 0.04 \\

2025-Dec-27.97 & FeReRo2 & 18.1 & 2.5
& \multicolumn{2}{c}{---}
& \multicolumn{2}{c}{---}
& 0.79 & 0.03
& \multicolumn{2}{c}{---} 
& \multicolumn{2}{c}{---} & \multicolumn{2}{c}{---} & \multicolumn{2}{c}{---} \\

2025-Dec-28.95 & FeReRo2 & 17.3 & 2.6
& \multicolumn{2}{c}{---}
& \multicolumn{2}{c}{---}
& 0.51 & 0.04
& \multicolumn{2}{c}{---} 
& \multicolumn{2}{c}{---} & \multicolumn{2}{c}{---} & \multicolumn{2}{c}{---} \\

2025-Dec-29.35 & FORS2 & 17.0 & 2.6
& \multicolumn{2}{c}{---}
& \multicolumn{2}{c}{---}
& 0.18 & 0.05
& \multicolumn{2}{c}{---} 
& \multicolumn{2}{c}{---} & \multicolumn{2}{c}{---} & \multicolumn{2}{c}{---} \\

2026-Jan-04.18 & FORS2 & 12.4 & 2.8
& -1.09 & 0.35
& -1.16 & 0.21
& -1.21 & 0.19
& -0.90 & 0.17 
& 0.07 & 0.23 & 0.05 & 0.29 & -0.17 & 0.15 \\

2026-Jan-10.31 & FORS2 & 7.8 & 3.0
& -2.59 & 0.10
& -2.57 & 0.09
& -2.44 & 0.08
& -2.43 & 0.08 
& -0.09 & 0.08 & -0.14 & 0.12 & -0.01 & 0.06 \\

2026-Jan-14.98 & ALFOSC & 4.6 & 3.1
& \multicolumn{2}{c}{---}
& \multicolumn{2}{c}{---}
& -2.97 & 0.35
& \multicolumn{2}{c}{---} 
& \multicolumn{2}{c}{---} & \multicolumn{2}{c}{---} & \multicolumn{2}{c}{---} \\

2026-Jan-15.98 & FeReRo2 & 4.0 & 3.1
& \multicolumn{2}{c}{---}
& \multicolumn{2}{c}{---}
& -2.83 & 0.05
& \multicolumn{2}{c}{---} 
& \multicolumn{2}{c}{---} & \multicolumn{2}{c}{---} & \multicolumn{2}{c}{---} \\

2026-Jan-17.95 & FeReRo2 & 2.7 & 3.2
& \multicolumn{2}{c}{---}
& \multicolumn{2}{c}{---}
& -2.54 & 0.05
& \multicolumn{2}{c}{---} 
& \multicolumn{2}{c}{---} & \multicolumn{2}{c}{---} & \multicolumn{2}{c}{---} \\

2026-Jan-18.23 & FORS2 & 2.6 & 3.2
& -1.52 & 0.11
& -1.93 & 0.09
& -2.23 & 0.08
& -2.55 & 0.10 
& 0.41 & 0.08 & 0.31 & 0.12 & 0.18 & 0.07 \\

2026-Jan-20.13 & FORS2 & 1.5 & 3.3
& -1.08 & 0.10
& -1.26 & 0.09
& -1.54 & 0.08
& -1.72 & 0.08 
& 0.28 & 0.07 & 0.29 & 0.12 & 0.1 & 0.06 \\

2026-Jan-20.77 & FeReRo2 & 1.2 & 3.3
& \multicolumn{2}{c}{---}
& \multicolumn{2}{c}{---}
& -1.34 & 0.07
& \multicolumn{2}{c}{---}
& \multicolumn{2}{c}{---} & \multicolumn{2}{c}{---} & \multicolumn{2}{c}{---} \\

2026-Jan-21.13 & FORS2 & 1.1 & 3.3
& -0.86 & 0.10
& -0.91 & 0.09
& -1.02 & 0.07
& -1.29 & 0.07 
& 0.09 & 0.07 & 0.29 & 0.12 & 0.1 & 0.06 \\ 

2026-Feb-07.05 & FORS2 & 6.4 & 3.9
& -2.66 & 0.14
& -2.57 & 0.13
& -2.71 & 0.11
& -2.61 & 0.13 
& 0.02 & 0.11 & 0.14 & 0.17 & -0.05 & 0.1 \\
\hline
\end{tabular}
\vspace{0.5cm}
\noindent \parbox{\linewidth}{
\vspace{1mm}\textbf{Note.} R-band column includes broadband r' filter for ALFOSC; broadband R\_SPECIAL filter for FORS2; and narrowband Rx filter for FeReRo2. Other waveband measurements are taken only by FORS2: B-band includes narrowband FILT\_485\_37+68 and broadband b\_HIGH filters; V-band includes broadband v\_HIGH filter; and I-band includes narrowband FILT\_834\_48+71 and broadband I\_BESS filters. See Table~\ref{tableLog} and Fig.~\ref{FigFilters} for more details. 
}

\end{table*}

\subsection{Polarisation as a function of phase angle}
\begin{figure*}
\begin{center}
\includegraphics[width=0.9\textwidth]{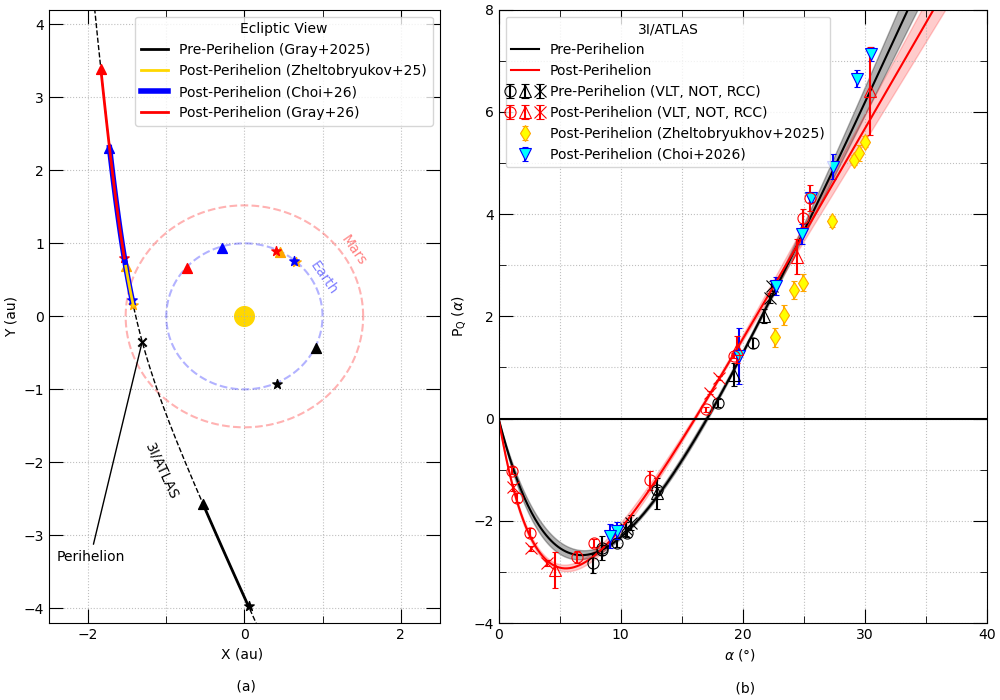}
\end{center}
\caption{\label{Figure1} (a) Orbital geometry of 3I projected on to the ecliptic plane (top-down view). The observation windows of \citet{gray2025ApJ}, \citet{zheltobryukhov2025}, \citet{choi2026}, and this work have been highlighted with solid-coloured lines on 3I's path. The beginning and end of the observing windows are indicated with solid stars and triangles, respectively, both on 3I's and Earth's paths. (b) R-band polarisation versus phase angle of 3I, plotted with data from the above-mentioned publications. The black and red lines represent the best-fit curves of the pre- and post-perihelion data (our data only) with $\pm 1\sigma$ uncertainty indicated by the shaded areas.} 
\end{figure*}

The numerical values of our aperture polarimetric measurements are listed in Table \ref{tablePQ}. The R-band measurements are plotted as a function of phase angle in Figure \ref{Figure1}b. For comparison, we include all previously published polarimetric measurements of 3I, which include pre-perihelion data by \citet{gray2025ApJ} obtained at heliocentric distances $r_H$=4--2.6 au, and phase angles $\alpha$=7.7--22.4$^{\circ}$, as well as post-perihelion data from \citet[$r_H$=1.4--1.6 au, $\alpha$=22.6--30$^{\circ}$]{zheltobryukhov2025} and \citet[$r_H$=1.5--2.9 au, $\alpha$=30.5--9.1$^{\circ}$]{choi2026}. To illustrate the timing and overlap in observing campaigns, we plot the orbital geometry of 3I relative to Earth and highlight the observing window of each dataset in Figure \ref{Figure1}a. 

We model the polarimetric phase curves of 3I using the following linear-exponential function \citep{muinonen2009}, with three free parameters ($A, B,$ and $C$) which shape the curve:
\begin{equation}
  \pr(\alpha) = A \left({\rm e}^{(-\alpha/B)} - 1\right) + C\,\alpha
\label{Eq_Fit}
\end{equation}
To ensure internal consistency and to allow for independent comparison, we fit the model to our data only. The resulting best-fit parameters are listed in Table \ref{tableCurve}.

Our new measurements show that the polarimetric phase curve of 3I remains largely unchanged following the perihelion passage, retaining the pronounced NPB reported previously in \citet{gray2025ApJ}. In fact, the R-band post-perihelion data indicate a slightly deeper NPB than the pre-perihelion data, with a larger (absolute) polarisation minimum value of $\pmin = -2.9 \%$ shifted to a smaller phase angle of $\amin = 5.5^{\circ}$. As highlighted in \citet{gray2025ApJ}, the $\pmin = -2.7\%$ and $\amin = 6.8^{\circ}$ values initially measured were effectively the upper limits only due to the lack of small phase angle coverage and the absence of the expected turnover toward zero polarisation. We therefore attribute the differences in the derived NPB parameters to the improved phase angle coverage in the post-perihelion dataset, rather than to intrinsic changes in the comet. 

At very small phase angles ($\alpha \leq 2^{\circ}$), \citet{gray2025ApJ} speculated that the polarimetric phase curve of 3I extrapolated to very small phase angles resembled that of certain small TNOs and Centaurs (see Fig.~1 of the article). However, due to the lack of data within this phase angle range in the pre-perihelion dataset, this finding was considered tentative. The addition of multiple observations at small phase angles in the post-perihelion dataset now confirms that the polarimetric behaviour of 3I matches closely that of small TNOs (diameters 310--670~km) Huya, Ixion, Varuna, and 1999 DE$_9$, as well as Centaur Pholus \citep{bagnulo2006A&A,belskaya2010}. 

In the PPB, 3I is characterised by a particularly high degree of linear polarisation. At $\alpha=30^{\circ}$, 3I is polarised approximately 3.5 percentage points (p.p.) more than typical high-polarisation SSC, and about 2 p.p. more than C/Hale-Bopp and 2I/Borisov. Thus, in both the NPB and PPB, the polarisation of 3I exceeds that of other comets. 

Compared to the other post-perihelion measurements, our data agree smoothly with those reported by \citet{choi2026}, but are systematically offset from those reported by \citet{zheltobryukhov2025}. This discrepancy is explored further in Section~\ref{sect_discussion}.

\begin{table*}
\caption{Polarimetric phase curve parameters of 3I according to the best fit model, including minimum value of polarisation (\pmin{}), phase angle of polarisation minimum (\amin), inversion angle (\ainv), and slope of curve at inversion (\textit{h}).}
\centering
\begin{tabular}{lcr@{$\pm$}lr@{$\pm$}lr@{$\pm$}lr@{$\pm$}l}
\hline \hline
    
Publication & Waveband & \multicolumn{2}{c}{\pmin ($\%$)} & \multicolumn{2}{c}{\amin ($^{\circ}$)} & \multicolumn{2}{c}{\ainv ($^{\circ}$)} & \multicolumn{2}{c}{$h$ ($\%$~deg$^{-1}$)} \\
\hline
Pre-perihelion & R & -2.67 & 0.08 & 6.79 & 1.34 & 17.08 & 0.12 & 0.416 & 0.008 \\
(Gray et al. 2025) \\
\hline 
Post-perihelion & B & -2.58 & 0.07 & 7.33 & 2.48 & 18.15 & 0.21 & 0.387 & 0.025 \\
(This work) & V & -2.67 & 0.06 & 6.68 & 0.97 & 17.27 & 0.16 & 0.393 & 0.013 \\ 
          & R & -2.94 & 0.03 & 5.52 & 0.30 & 16.10 & 0.06 & 0.392 & 0.008 \\ 
          & I & -2.89 & 0.05 & 5.02 & 0.31 & 14.88 & 0.13 & 0.408 & 0.008 \\
\hline
\end{tabular}
\label{tableCurve}
\end{table*}

\subsection{Polarisation as a function of wavelength}
\label{Sect_PolCol}
\begin{figure*}
\begin{center}
\includegraphics[width=\textwidth]{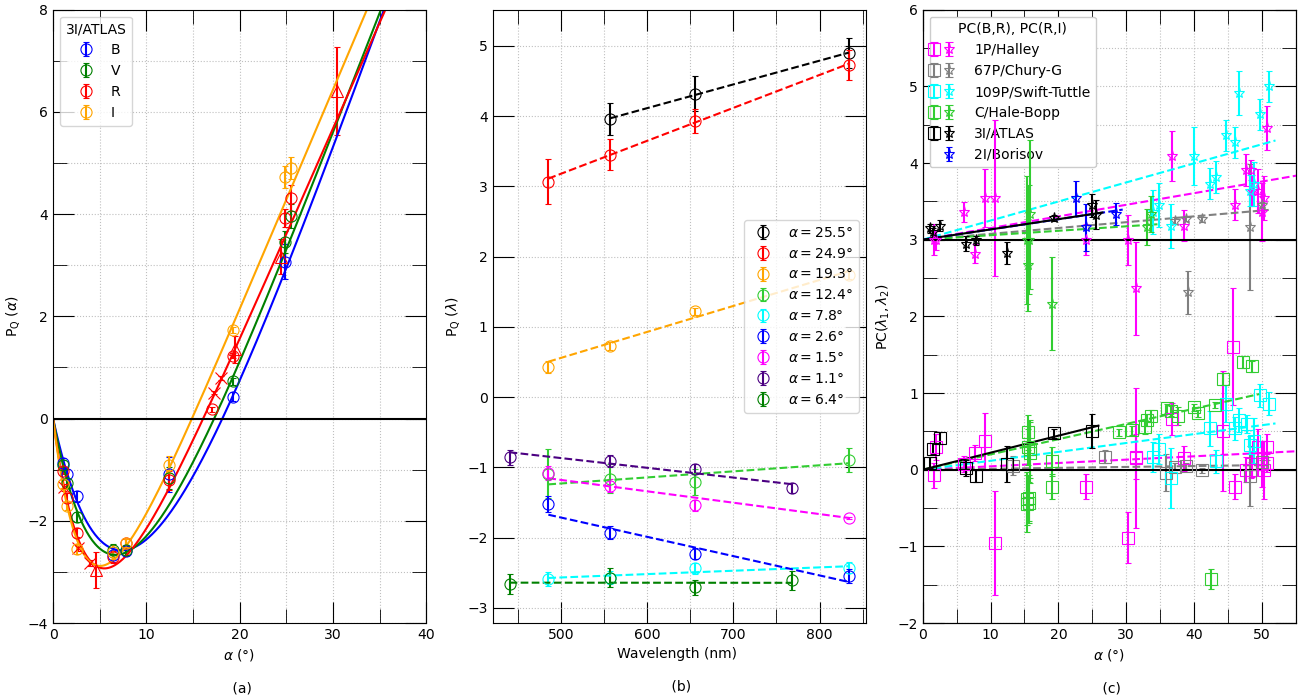}
\end{center}
\caption{\label{Figure2} (a) BVRI-band polarisation versus phase angle of 3I (new dataset only). (b) Polarisation versus wavelength at each phase angle. (c) Polarimetric colour (PC) plotted as a function of phase angle of 3I compared to that of other comets (see Appendix for details). The solid/dashed trends represent a weighted linear fit, i.e., $\Delta \rm{PC}(\lambda_1, \lambda_2)/\Delta \alpha$, fixed at zero. For display purposes, PC(B,R) (squares) and PC(R,I) (stars) are vertically offset.} 
\end{figure*}

The multi-band measurements and corresponding modelled polarimetric phase curves are plotted in Figure~\ref{Figure2}a. Our measurements show that 3I exhibits a pronounced NPB in all observed wavebands and a clear wavelength dependence is evident. Both the inversion angle and phase angle of minimum polarisation decrease almost linearly with increasing wavelength, while the slope parameter $h = \frac{dP_r}{d\alpha}|_{\alpha=\ainv}$ remains mostly constant across wavelengths. These trends indicate that the NPB becomes deeper and shifted to smaller phase angles with increasing wavelength. Consequently, the phase curves at different wavelengths are not simply offset versions of one another, but intersect within the NPB, implying that the wavelength dependence of polarisation changes across different phase angle regimes. 

In panel (b), we plot the degree of linear polarisation as a function of wavelength for each observing epoch. In the positive branch, the degree of polarisation increases approximately linearly with wavelength, which is consistent with that measured in optical bands by \citet{choi2026} who found a monotonic increase in polarisation between wavelengths 600--1200~nm. In contrast, the wavelength dependence of polarisation in the negative branch is less strictly linear. At intermediate phase angles within the NPB ($7<\alpha<15^{\circ})$, the polarisation tends to become less negative with increasing wavelength (i.e., the absolute value of polarisation decreases with wavelength), implying a reversal of the polarimetric spectral gradient relative to the PPB. At the smallest phase angles, however, the opposite trend is observed, with the polarisation becoming more negative towards longer wavelengths (i.e., the absolute value of polarisation increases). Thus, the spectral dependence of polarisation appears to change not only between the positive and negative branches, but also within the NPB itself. The detailed origin of this behaviour remains beyond the scope of this work, but the extensive phase angle coverage of 3I provides an opportunity to investigate the transition of the polarimetric spectral dependence across the NPB of solar system bodies, continuing works such as \citet{kiselev2005,belskaya2009} and \citet{bagnulo2015}.

To place this wavelength-dependent behaviour in the broader context of cometary polarimetry, we next compare the polarimetric colour (PC) of 3I with that of other solar system comets in panel (c). We define PC as the change in absolute polarisation with wavelength. If $P_r(\lambda_X)$ is the polarisation measured in a given filter centred at wavelength $\lambda_X$, we calculate PC as:
\begin{equation}
\rm{PC}(\lambda_1, \lambda_2) = \frac{d|P_r(\lambda)|}{d\lambda} = 
    \frac{|P_r(\lambda_2)| - |P_r(\lambda_1)|}{\lambda_2 - \lambda_1}
\end{equation}
According to this definition, positive (red) values indicate that the polarisation increases with wavelength, while negative (blue) values indicate the opposite behaviour. The calculated PC values for pairs (B,R), (V,R), and (I,R) are listed in Table~\ref{tablePQ}. As mentioned in Section \ref{Sect_Observations}, the B- and I-bands correspond to narrowband filters in all FORS2 epochs except for the final two. These differences in wavelength have been accounted for when calculating the PC. 


The derived PC quantities show generally small and positive values, indicating a predominantly red PC, consistent with typical SSCs. Figure~\ref{Figure2}c compares the PC(B,R) and PC(R,I) values of 3I plotted as a function of phase angle with those of other comets (see Appendix for details). We do not include a comparison of PC(V,R) due to the limited availability of such measurements in other comets. For each comet, we calculate the approximate rate of change of PC with phase angle, i.e. $\Delta \rm{PC}(\lambda_1,\lambda_2)/\Delta \alpha$, using weighted linear fits, where each measurement is weighted by the inverse square of its uncertainty. A similar approach was adopted by \citet{belskaya2009} to study the wavelength dependence of polarisation of asteroids. These trends indicate that, despite significant scatter, SSCs typically show a red PC that becomes more pronounced with phase angle.

Our results confirm that 3I follows the same overall trend. Its PC remains mostly positive and shows a modest increase with phase angle, comparable to that of other comets, including the polarimetrically unique cases of C/Hale-Bopp and 2I/Borisov, suggesting some commonalities in dust properties despite 3I's interstellar origin. However, given the limited number of comets with contemporaneous multi-band observations, together with the systematic differences in filter sets between studies, it is difficult to robustly quantify general trends in the phase angle dependence of PC, and thus, contextualise the PC measurements of 3I. Therefore, the comparison presented here should be regarded as illustrative rather than quantitative.

\subsection{Coma morphology}

\begin{figure*}
\begin{center}
\includegraphics[width=\textwidth]{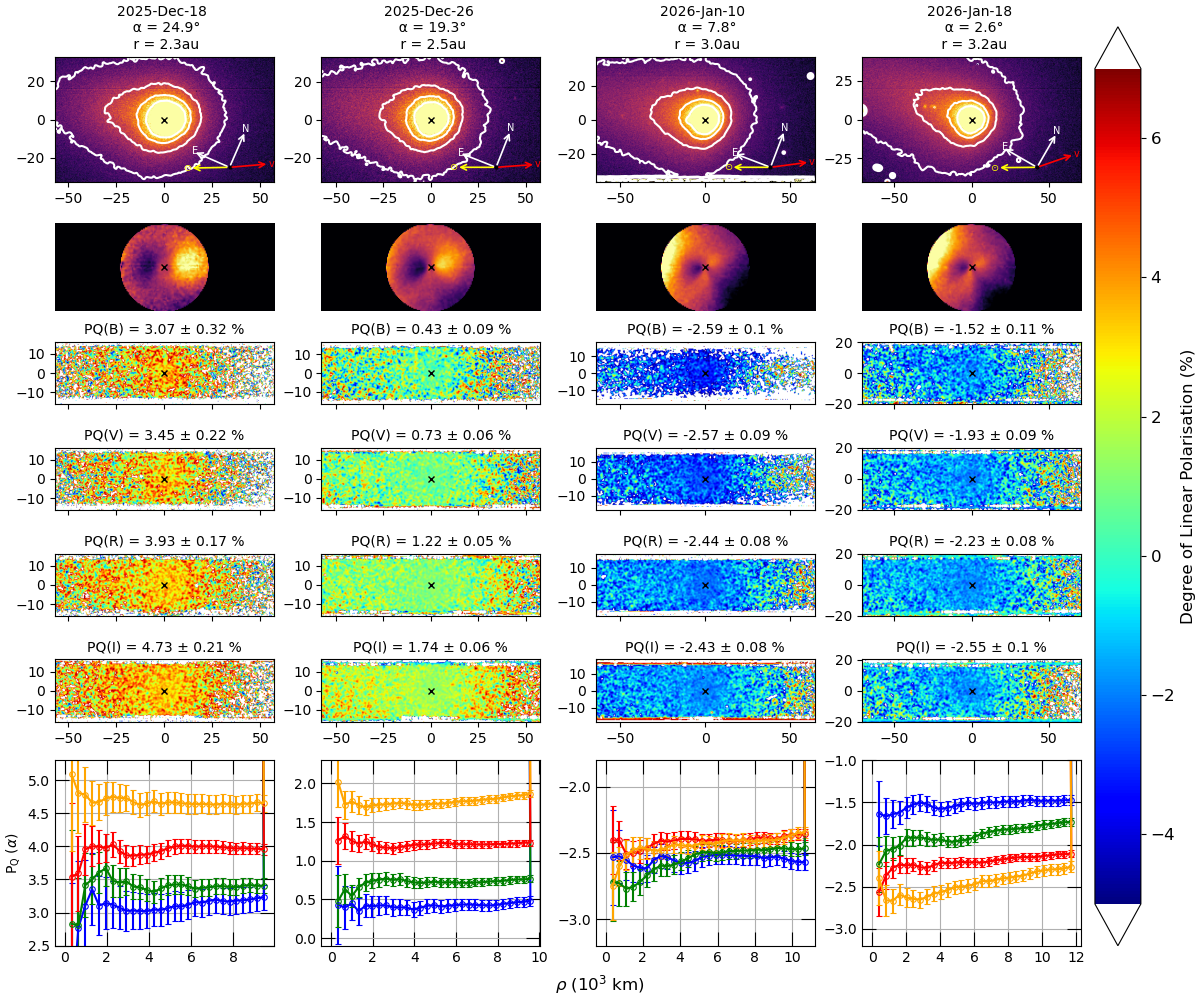}
\end{center}
\caption{\label{Figure3} Subset of imaging polarimetric data of 3I. \textit{First row:} R-band imaging maps with isophotes. The arrows indicate the celestial north and east directions, the projected solar direction and projected heliocentric velocity direction. X marks the comet photocentre. \textit{Second row:} R-band imaging maps after division by azimuthal average. \textit{Third to sixth rows}: BVRI polarimetric maps. The colour of each pixel represents the value of polarisation, as shown in the scale to the right. Pixels with values outside this colour scale are given in white and are considered noise. \textit{Seventh row:} BVRI polarisation (\pq) measured as a function of aperture radius, increasing in one-pixel increments. The waveband-related colour scheme follows that of Fig~\ref{Figure2}a.} 
\end{figure*}

Figure \ref{Figure3} presents a subset of our FORS2 imaging and polarimetric maps, together with the degree of linear polarisation measured as a function of aperture radius in the final row. Each column corresponds to a single observing epoch. 

The first row shows the acquisition images obtained immediately prior to the polarimetric observations, overlaid with isophotal contours to highlight the morphology of the coma and tail. Although deep imaging maps can be constructed with FORS2 imaging polarimetric data by co-adding the individual polarimetric images as done in \citet{gray2025ApJ}, the effective field of view (FoV) is restricted by the Wollaston-mask optics, resulting in vignetted strips of approximately 22'' width. Since the coma of 3I extends well beyond this spatial coverage, we display the acquisition images to provide a more expansive view of the coma and tail morphology. 

Compared to the pre-perihelion observations presented in \citet{gray2025ApJ}, 3I exhibits a substantially more extended coma and more prominent tail in the post-perihelion observations. The innermost isophotes show slight departures from circular symmetry, with an apparent enhancement in the sunward direction and an additional asymmetry towards the south. Such structures may indicate localised regions of enhanced dust emission from the nucleus, although projection effects and the superposition of different dust populations along the line of sight may also contribute to the observed morphology. At larger distances from the photocentre, the outer isophotes trace an extended coma and tail predominantly oriented in the solar direction. The absence of a prominent anti-solar tail is consistent with several independent studies, which conclude that the coma is dominated by relatively large dust grains that are only weakly affected by solar radiation pressure \citep[e.g.,][]{jewitt2025,chandler2026,ren2026}. A similarly sunward extended dust tail has also been reported in post-perihelion observations by \citet{moreno2026}, \citet{lisse2026}, \citet{belyakov2026arXiv_dust}, and \citet{roth2026}.

To investigate the coma morphology in greater detail, we applied several image-enhancement techniques to the R-band deep imaging maps using the web-based Cometary Coma Image Enhancement Facility\footnote{\url{https://cie.psi.edu/}} \citep{samarasinha2013}. We tested division by the azimuthal average, division by azimuthal median, and division by $1/\rho$ profile. We found that these different approaches produced broadly similar morphological structures, suggesting that the main features identified are not strongly dependent on the particular enhancement method. We therefore show the images obtained after division by azimuthal average in the second row of Figure~\ref{Figure3}. 

The enhanced images reveal some low-contrast jet-like enhancements, although their physical nature cannot be established unambiguously. In the December observations, a feature extending approximately in the anti-sunward direction is visible, together with a fainter structure towards the south. In the January observations, the southern feature appears more pronounced, while the anti-sunward structure appears to shift towards the northern direction. In pre-perihelion observations, \citet{serraricart2026} identified a faint jet whose position angle oscillated around $\sim280^{\circ}$ with a periodicity consistent with the comet's rotation period. In contrast, \citet{ren2026} found no convincing evidence for jets in observations obtained close to perihelion. Thus, the possible re-emergence of jet-like features may reflect variable dust emission from localised active regions, changing viewing geometry, nucleus rotation or projection effects. Distinguishing between these possibilities requires a more detailed analysis and remains beyond the scope of this work.

\subsection{Spatially resolved polarisation}

The next four rows of Figure~\ref{Figure3} display the corresponding BVRI polarimetric maps. A Gaussian filter has been applied to reduce noise while preserving the overall spatial structure of the coma. Compared to the pre-perihelion maps in \citet{gray2025ApJ}, the post-perihelion observations reveal a substantially more extended region of measurable polarisation, consistent with the much larger spatial extent of the coma at these epochs. Despite the possible morphological structures found in the imaging and enhanced images, however, polarimetric maps show no corresponding strong spatial variations. Instead, the polarisation is generally smooth and homogeneous across both the coma and visible tail, with no clear polarimetric signatures associated with the dust tail or possible jet-like structures. A weak gradient in polarisation towards the outer coma is visible in some of the maps. However, this occurs primarily in regions where the coma becomes increasingly diffuse and the signal-to-noise ratio decreases, and may therefore result from increasing measurement uncertainties rather than a genuine spatial variation. 

The homogeneity of the polarisation distribution is also reflected in the aperture polarimetry. The final row of Figure \ref{Figure3} shows the measured polarisation as a function of aperture radius, with the aperture increasing in increments of one pixel. Aside from small-scale variations at the smallest aperture sizes, likely resulting from small-number statistics of the enclosed pixels, the measured polarisation rapidly reaches a plateau and remains approximately constant within uncertainty with increasing aperture size. 

This overall homogeneous distribution is consistent with both the pre-perihelion polarimetric maps presented in \citet{gray2025ApJ} and post-perihelion aperture polarimetry measurements of \citet[][see their Fig.~A1]{choi2026}, further indicating that the integrated polarisation signal is not strongly dependent on the spatial extent of the coma within our field of view. 

Taken together, our imaging and polarimetric observations reveal an interesting contrast between the morphological evolution and polarimetric stability of 3I. This suggests that the dust associated with the low-contrast morphological features either has broadly similar optical properties to the bulk of the coma dust, or that any differences are below the sensitivity of our observations. More generally, the absence of strong spatially resolved polarimetric variations supports the interpretation that the unusual polarimetric phase curve of 3I is characteristic of its dominant dust population rather than being driven by a small, spatially confined component of the coma.

\section{Discussion}
\label{sect_discussion}

\subsection{Stability of the polarimetric phase curve}
The main result of our post-perihelion observations is that the polarimetric phase curve of 3I remains largely unchanged following its perihelion passage. Likewise, post-perihelion polarimetric observations by \citet{choi2026} found no significant change in the polarimetric phase curve compared to pre-perihelion. Similar results were reported by \citet{ivanova2026ACM} and \citet{wohler2026ACM}. 


These findings differ from those presented by \citet{zheltobryukhov2025}, who report systematically lower post-perihelion polarisation values in the positive branch. Based on dedicated light-scattering modelling, \citet{zubko2025} suggested that the unusual pre-perihelion polarimetric phase curve in \citet{gray2025ApJ} can be reproduced only if the coma contains a substantial proportion of pure water-ice particles. More specifically, their model suggests that 30--40\% of particles populating the inner pre-perihelion coma consist of pure water-ice, while the remaining particles are composed primarily of Mg-rich silicates and organics/amorphous carbon. Applying the same modelling framework to their post-perihelion observations, \citet{zheltobryukhov2025} find that the systematically lower polarisation can be reproduced by removing the water-ice component from their previously modelled pre-perihelion coma. Thus, the authors attribute the lower post-perihelion polarisation to the loss of this icy component as the comet moved to smaller heliocentric distances. In this interpretation, the pre-perihelion observations ($r_H\geq2.6$~au) sample the comet beyond the snowline, where icy grains can survive within the coma without sublimating, whereas their post-perihelion observations ($r_H=1.4-1.6$~au) probe a warmer solar environment in which water-ice grains are expected to sublimate rapidly. 

However, considering the persistence of the polarimetric phase curve observed in both our data and those of \citet{choi2026}, it is difficult to reconcile it with a scenario in which the dominant scattering particle population changes substantially across perihelion. Notably, the post-perihelion dataset of \citet{choi2026} begins at $r_H=1.5$~au, where the effects of water-ice sublimation should already be well established under the interpretation of \citet{zheltobryukhov2025}. Likewise, the discrepancy is unlikely to arise from differences in the adopted aperture size, as the aperture radius used by \citet{zheltobryukhov2025} ($\sim$10,000~km) is between that used in our observations (2,000--6,000~km) and by \citet{choi2026} (15,000--20,000~km). Instead, the smooth continuity between the pre- and post-perihelion datasets suggests that the bulk optical scattering properties of the coma dust remained mostly stable throughout the apparition. In the following section, we explore alternative interpretations of the observed polarimetric behaviour through additional light-scattering modelling.

\subsection{Stability of the dust properties}

Although the activity of 3I evolved substantially throughout its passage through the inner Solar System, several independent groups indicate that the optical properties of its dust remained remarkably stable. In this context, the persistence of the polarimetric phase curve found in this work is not an isolated result, but rather forms part of a broader observational picture. 

Visible spectroscopic observations consistently report a moderately red dust continuum, with optical spectral slopes of approximately $15-20\%$ per 100~nm throughout both the pre- and post-perihelion phases of the apparition \citep{delaFuenteMarcos2025AA,hoogendam2026_spectrophot,lazzarin2026,medler2026,opitom2025,seligman2025,yang2025,kawakita2026,bockelee2026}. Although the exact spectral slope sometimes depends on the wavelength interval over which it is measured, the overall spectral shape remains largely unchanged across perihelion. Similarly, long-term photometric monitoring found no significant evolution in the broadband colours of the dust coma as a function of heliocentric distance or aperture size \citep{hoogendam2026_spectrophot, gillian2026arxiv,moreno2026}. These studies consistently show that 3I maintained a continuum dominated by moderately red dust despite significant changes in activity. 

In contrast, observations of the volatile component reveal pronounced perihelion asymmetries. Production rates of CN, Ni, Fe, and other species evolved substantially throughout the apparition \citep{medler2026,zhao2026,hutsemekers2026}, while the dominant volatile source transitioned from a CO$_2$-rich coma at large heliocentric distances to H$_2$O-dominated activity as it approached perihelion \citep{cordiner2025,lisse2026,combi2026,tan2026,roth2026}. Estimates of the cumulative mass loss further suggest that meters of surface material may have been removed during the apparition \citep{frincke2026arXiv,belyakov2026arXiv_dust}. Despite these significant changes in activity and volatile composition, neither the optical colours nor the polarimetric phase curve exhibit strong evolution. 

Taken together, these results suggest that the bulk optical properties of the dust population remained comparatively stable throughout the apparition. This favours the interpretation in which the observed polarimetric behaviour primarily reflects the intrinsic scattering properties of the refractory dust population rather than transient changes associated with volatile sublimation.

\section{Modelling}

The polarimetric phase curve of 3I exhibits an unusual negative polarisation branch, while the internal structure and dust composition of interstellar comets are generally unknown. We therefore conducted a preliminary parameter-space study of dust models spanning a range of material compositions and porosities. Three aggregate morphologies were considered: highly porous ballistic cluster-cluster aggregates (BCCAs), moderately porous ballistic particle-cluster aggregates (BPCAs), and dense ballistic aggregates with two migrations (BAM2). The aggregates consisted of spherical monomers of equal radius, $a = 130$\,nm. The material compositions considered were amorphous iron-poor pyroxene, amorphous iron-rich olivine, weakly and heavily processed organic matter, water ice, and tholins. The refractive indices used at a wavelength of $\lambda = 660$\,nm are listed in Table \ref{table:refind}.

\begin{table*}[t]
\caption{Refractive indices of materials.}
\label{table:refind}
\centering
\begin{tabular}{l c c}
\hline\hline
Material & Refractive index & Reference\\
\hline
iron-poor pyroxene & $m = 1.58 + i3.318\times 10^{-4}$ & \cite{Dorschner1995} \\
iron-rich olivine & $m = 1.79 + i7.3 \times 10^{-2}$ & \cite{Dorschner1995} \\
unprocessed organics & $m = 1.57 + i8.5 \times 10^{-2}$ & \cite{jenniskens1993}\\
processed organics &  $m = 1.94 + i3.6 \times 10^{-1}$ & \cite{jenniskens1993}\\
water ice &  $m = 1.31 + i1.66 \times 10^{-8}$& \cite{Warren1984}\\
tholins & $m = 1.56 + i9 \times 10^{-4}$ & \cite{Ramirez2002}\\
\hline
\end{tabular}
\end{table*}

The computations were carried out using the fast superposition T-matrix method (FaSTMM2) \citep{markkanen2017, markkanen2026}. FaSTMM2 computes the full Mueller matrix for a given aggregate geometry and composition. Four aggregate sizes, consisting of 1024, 4096, 16384, and 65536 monomers, were considered. The edge length of the smallest bounding box enclosing the aggregate ranged from 13.5 to 190\,µm for BCCA, 5.6 to 22\,µm for BPCA, and 4.8 to 18\,µm for BAM2. For each case, the results were averaged over 10 aggregate realisations and 16 orientations to  obtain the ensemble-averaged Mueller matrix. These computations extend the dataset that was previously used to model 3I colour evolution for the inbound phase by \cite{Santana2026}. Finally, the parametrisation given by Eq. (\ref{Eq_Fit}) was fitted to the computed degree of linear polarisation, allowing the parameters $P_{\rm min}, \alpha_{\rm min}, \alpha_{0}$, and $h$ to be determined. 

\begin{figure}
\begin{center}
\includegraphics[width=0.5\textwidth]{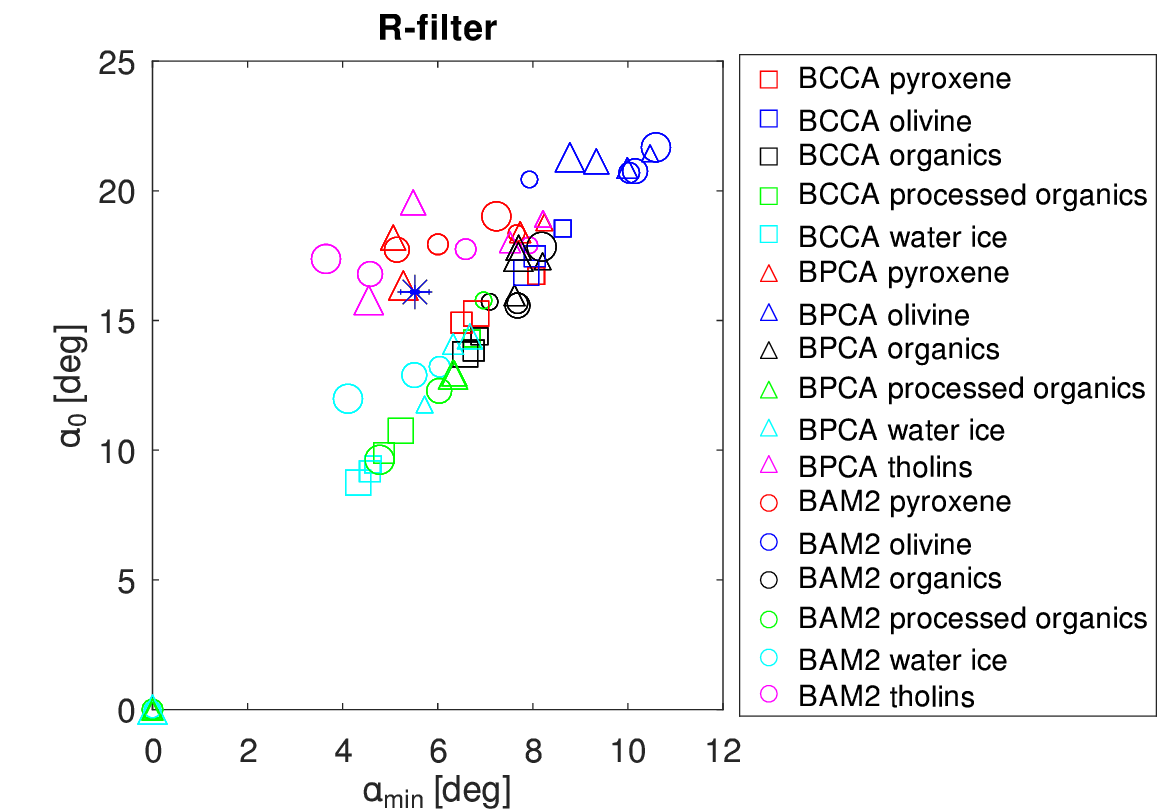}
\end{center}
\caption{\label{fig:model} Inversion angle as a function of the minimum polarisation angle. Different materials are denoted by different colours whereas aggregate morphologies are indicated by different markers. Marker size indicates the aggregate size. The smallest marker corresponds to 1024 monomers whereas the largest to 65536 monomers. The blue asterisk is the post perihelion value from Table \ref{tableCurve}. Zero values indicate the absence of the negative polarization branch.} 
\end{figure}

Figure \ref{fig:model} shows the inversion angle $\alpha_0$ with respect to the minimum polarisation angle $\alpha_{\rm min}$ for aggregates with different morphologies, compositions, and sizes. Interestingly, all moderately or strongly absorbing materials follow the linear relation $\alpha_0 = 2\alpha_{ \rm min}$ indicating that the negative polarisation branch is symmetric. In contrast, the weakly absorbing materials such as iron-poor pyroxene, tholins, and water ice deviate from this trend, resulting in an asymmetric negative polarisation branch. This suggests that 3I may contain a larger fraction of weakly absorbing material than typical SSCs, which exhibit a symmetric negative polarisation branch as demonstrated in Figure~\ref{fig:model}.

\begin{figure}
\begin{center}
\includegraphics[width=0.5\textwidth]{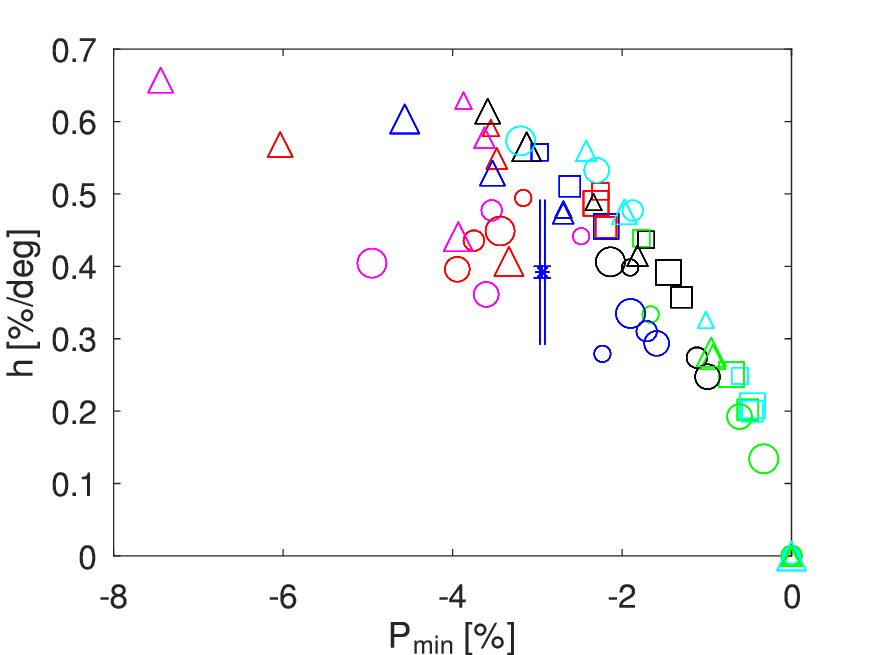}
\end{center}
\caption{\label{fig:model2} Slope of the polarisation curve at the inversion angle with respect to the minimum polarization. The markers and colours are defined in Fig. \ref{fig:model}.} 
\end{figure}

Figure \ref{fig:model2} plots the slope $h$ at the inversion angle with respect to the minimum polarisation $P_{\rm min}$. This plot suggests that to produce $P_{\rm min} < -3\%$ at $h \approx 0.4\%/\text{deg}$,  3I's dust should contain weakly absorbing materials such as tholins or iron-poor silicates, possibly linearly mixed with a more strongly absorbing component. Several combinations of materials and aggregate morphologies reproduce the observed polarimetric data when the size distribution is treated as a free parameter. However, all such models require moderately porous BPCA or dense BAM2 aggregates composed of weakly absorbing materials. Including other wavelengths and dust dynamics, we can further constrain the dust model.

Dynamical dust modelling suggests that the minimum particle size should be approximately 10\,µm \citep{moreno2026}. This is supported by the absence of an anti-solar tail as seen in Figure~\ref{Figure3}. Smaller particles would be more affected by solar radiation pressure, forming a bright anti-solar tail. To compute scattering by particles substantially larger than 10\,µm, we employed the static-structure-factor-corrected radiative transfer-coherent backscattering method (RT-CB) \citep{markkanen2023}. RT-CB is an approximate numerical method that enables the computation of scattering by very large clusters composed of equal-sized spherical monomers. The monomers must be identical in size and composition but may possess a core–mantle structure. RT-CB has been validated against exact computational methods for both sparsely and densely packed particulate media \citep{muinonen2012,muinonen2025}.

Our goal was to find a model that simultaneously reproduces the degree of linear polarisation in R- and B-filters, and the observed B-R colour. \cite{Santana2026} modelled pre-perihelion R-band polarisation and colours using a subset of particles presented in Figure~\ref{fig:model}. Here, we increase the particle size range using RT-CB modelling, and include post-perihelion data as well as the B-band polarisation to further constrain the model. Because the colour did not change significantly after the perihelion \cite{moreno2026}, we used B-R colour of 1.4 mag as derived for the inbound phase by \cite{Santana2026}.  

Using RT-CB, we performed a parameter-space study by varying monomer compositions and sizes to find an acceptable fit. The results of our RT-CB computations are summarised as follows: The shape of the negative polarisation branch requires particles with low absorption, while the slightly red polarimetric colour indicates monomers with radii of approximately $0.1$\,µm. Larger monomers produce a greater difference between the modelled polarisation phase curves in the B and R filters. A red B–R colour combined with 0.1\,µm-sized monomers requires a material composition with a significant decrease in the imaginary part of the refractive index across visible wavelengths, such as tholins.   

\begin{figure}
\begin{center}
\includegraphics[width=0.5\textwidth]{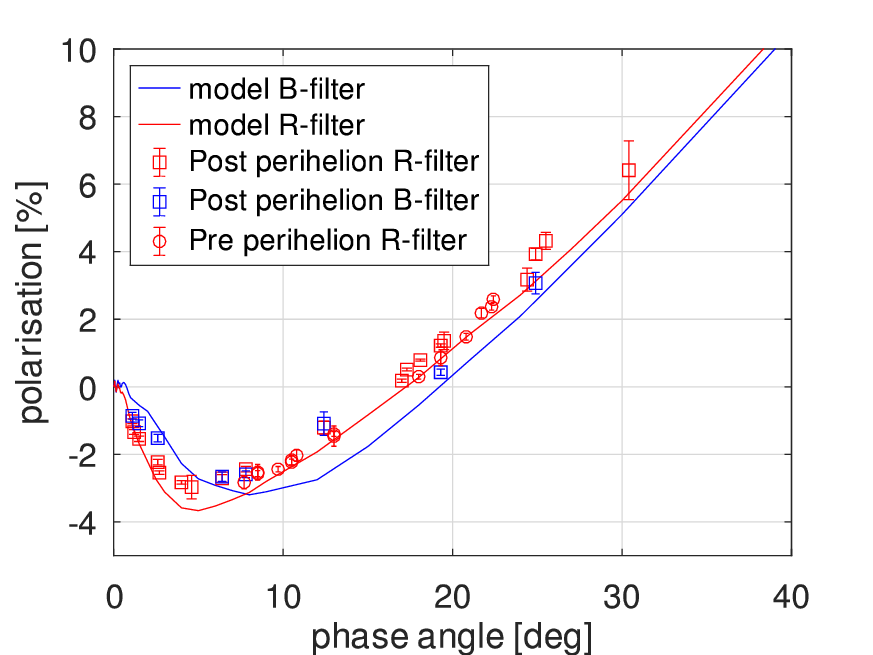}
\end{center}
\caption{\label{fig:model3} Modelled polarisation phase curve for aggregates larger than 10\,µm consisting of silicate core tholin mantle monomers using RT-CB.} 
\end{figure}

Figure \ref{fig:model3} shows a computed polarisation phase curve using RT-CB that provides an acceptable fit to the observations. The monomers consisted of an iron-poor silicate core with a radius of 0.08\,µm, and a tholin mantle with a thickness of 0.02\,µm. The porosity of the entire particle was 65\%, and the stickiness parameter in the RT-CB solver was 0.11. Low porosity and small stickiness are required to reach a red colour. The differential power-law index of the particle size distribution was -3.5 with minimum and maximum radii of 10\,µm and 1000\,µm, respectively. Particle size has only a minor effect on the polarisation and colour within the considered range. Increasing the power-law index results in only a slight decrease in polarisation and a redder colour. Therefore, for these particles, dynamical sorting does not produce substantial spatial variations in the coma, consistent with the observed polarimetric maps in Figure~\ref{Figure3}.       

Overall, our modelling results suggest that the dominant contribution to the scattered light at visible wavelengths originates from moderately porous weakly absorbing aggregates composed of 0.1\,µm monomers made of iron-poor silicates and tholin-like organic matter. Water ice particles, as a weakly absorbing component, are not sufficient because of their blue colour and too small $\alpha_{0}$. This contrasts with typical solar system comets, which often exhibit a stronger contribution from highly absorbing particles, such as those observed in comet 67P \citep{rosenbush2026}. Consequently, the geometric albedo of the 3I dust particles predicted by our model is approximately 0.7, which is an order of magnitude higher than that estimated for 67P. Such bright particles produce a significant opposition surge, which was indeed observed by \cite{moreno2026} and \cite{hui2026}. It should be noted, however, that the coma may still contain a significant fraction of typical cometary particles in mass, but their contribution to the total scattered light is small because of their significantly lower brightness.      

\section{Conclusions}

We present an extensive post-perihelion polarimetric study of the third discovered interstellar object, 3I/ATLAS (C/2025 N1), combining new observations obtained with VLT/FORS2, NOT/ALFOSC, and RCC/FoReRo2. Together with the pre-perihelion dataset presented by \citet{gray2025ApJ}, these observations provide the most comprehensive polarimetric phase-angle coverage obtained for an interstellar object to date.

Our principal observational findings are as follows:

\begin{itemize}
    \item The post-perihelion measurements connect smoothly with the pre-perihelion polarimetric phase curve, confirming that 3I is characterised by an unusually deep and narrow negative polarisation branch together with a comparatively small inversion angle. Although the post-perihelion dataset yields a slightly larger absolute value of polarisation minimum and a slightly smaller phase angle of minimum polarisation, these differences are most naturally explained by the improved coverage at very small phase angles.
    \item Multi-band observations show that 3I exhibits a red polarimetric colour that increases with phase angle in a manner broadly consistent with SSCs, despite the unusual shape of its polarimetric phase curve.
    \item Imaging data reveal a substantially more extended dust coma compared to the pre-perihelion campaign, with sunward-pointing dust tail and evidence for low-contrast features. 
    \item Imaging polarimetric maps show a smooth and spatially homogeneous degree of polarisation across the coma and tail, and aperture polarimetry demonstrates that the measured polarisation is largely independent of aperture size over a broad range of projected cometocentric distances. 
    \item The observed polarimetric phase curves, including the unusual shape of the negative polarisation branch and the red polarimetric colour, are best reproduced by models containing weakly absorbing moderately porous aggregates larger than 10\,µm composed of submicron monomers. The modelled particles have a high geometric albedo, substantially exceeding that of typical solar system cometary dust. 
\end{itemize}

The apparent stability of the polarimetric behaviour is noteworthy given the substantial evolution in the comet's activity, volatile production, and morphology reported by numerous independent studies. However, the persistence of the polarimetric phase curve found in this work is not an isolated result. Rather, it complements a growing body of evidence that the optical properties of 3I's dust remain comparatively stable throughout its apparition, with several spectroscopic and photometric studies reporting a consistently red continuum across perihelion. Together, these results suggest that the bulk optical scattering properties of the coma dust remained stable throughout apparition. 

Overall, the observations of 3I/ATLAS provide an additional example of the diversity of both ISOs and SSCs. The extensive characterisation of 3I throughout its passage through the Solar System will provide an important benchmark for studies of future interstellar visitors and for placing their properties in the broader context of small-body populations within the Solar System. 

\vspace{5mm}

\section*{Acknowledgements}

Based on observations obtained with the FORS2 instrument at the ESO Telescopes at the La Silla Paranal Observatory under program ID 116.2APG.001 (PI Gray). All FORS2 raw data and calibrations are available at {\tt archive.eso.org}.

Based on observations made with the Nordic Optical Telescope, owned in collaboration by the University of Turku and Aarhus University, and operated jointly by Aarhus University, the University of Turku and the University of Oslo, representing Denmark, Finland and Norway, the University of Iceland and Stockholm University at the Observatorio del Roque de los Muchachos, La Palma, Spain, of the Instituto de Astrofisica de Canarias. The NOT data were obtained under program ID P72-408 (PI Gray). The data presented here were obtained in part with ALFOSC, which is provided by the Instituto de Astrofisica de Andalucia (IAA) under a joint agreement with the University of Copenhagen and NOT. 

Based on observations obtained at BNAO Rozhen during semester 2025A under the program "Investigation of the physico-chemical properties of different classes of comets" led by G. Borisov.

BNAO Rozhen is a facility of the National Roadmap for Research Infrastructure 2020-2027 (RACIO project), financially coordinated by the Ministry of Education and Science of the Republic of Bulgaria.

G.B.\ acknowledges partial support from grant K$\Pi$-06-H88/5 by the Bulgarian National Science Fund.

Z.G., A.P., and K.M. acknowledge support by the Research Council of Finland (grants \#359893 and \#336546)

J.M. acknowledge support from the German Research Foundation (DFG) grant no. 517146316. This work used the supercomputer Phoenix and was supported by the Gauß-IT-Zentrum of the University of Braunschweig (GITZ)

R.C.D. and M.G. acknowledge support from the Research Council of Finland (grant \#361233).

\bibliography{References}{}
\bibliographystyle{aasjournal}

\newpage
\appendix 

\section{Instrumental setup and observing strategy}

The post-perihelion campaign employed the same observing strategy, calibration procedures, and data reduction methods as described by \citet{gray2025ApJ}. Therefore, only the aspects that differ from the pre-perihelion campaign are summarised here.

\begin{figure*}
\begin{center}
\includegraphics[width=0.6\textwidth]{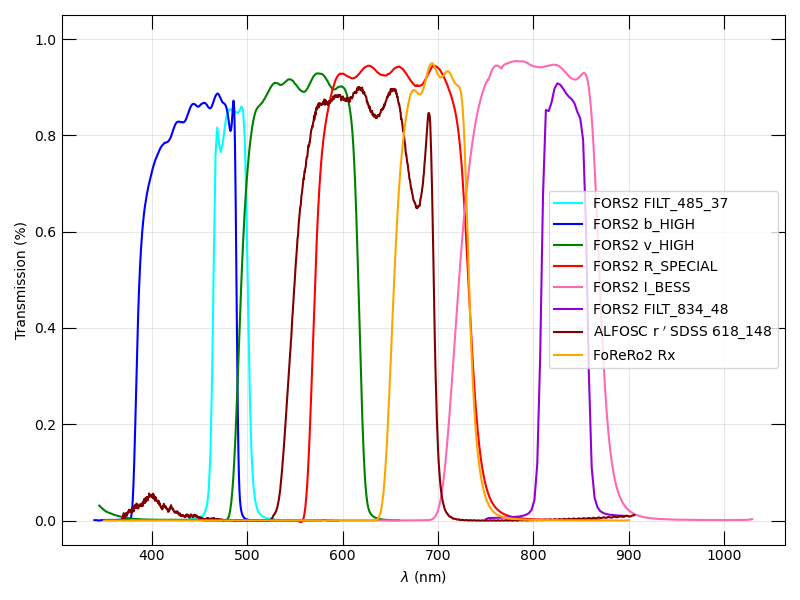}
\end{center}
\caption{\label{FigFilters} Transmission curves of filters used for this study. We describe the filters as follows: (Filter name, central wavelength, Full Width Half Maximum). FORS2: (b\_HIGH,~440~nm,~103.5~nm); (FILT\_485\_37,~485~nm,~37~nm); (v\_HIGH,~557~nm,~123.5~nm); (R\_SPECIAL,~655~nm,~165~nm); (I\_BESS,~768~nm,~138~nm); (FILT\_834\_48,~834~nm,~48~nm). ALFOSC: (r'\_SDSS 618\_148,~618~nm,~148~nm). FoReRo2: (Rx, 694~nm, 79~nm). }
\end{figure*}

Imaging polarimetric observations were obtained with FORS2 at the VLT, FoReRo2 at the RCC, and ALFOSC at the NOT. The transmission curves of all filters used are shown in Figure~\ref{FigFilters}, while their central wavelengths ($\lambda_{\rm c}$) and full widths at half maximum (FWHM) are given in the figure caption.

The FORS2 and FoReRo2 observations were carried out using the same instrumental configurations as in \citet{gray2025ApJ}. The only significant change concerned the ALFOSC observations. While the pre-perihelion campaign employed the FAPOL polarimetric mode, the post-perihelion observations were obtained using the Wedged Double Wollaston (WeDoWo) configuration \citep{oliva1997}. WeDoWo simultaneously produces four polarised images corresponding to position angles of $0^\circ$, $45^\circ$, $90^\circ$, and $135^\circ$, allowing the reduced Stokes parameters to be determined directly as: $\pq = Q/I =  (I_{0}-I_{90})/(I_{0}+I_{90})$ and $\pu = U/I = (I_{45}-I_{135})/(I_{45}+I_{135})$, where $I_\theta$ denotes the intensity measured at polarisation angle $\theta$ \citep{shurcliff1962}. 

The change in the ALFOSC configuration was motivated by the substantially more extended coma observed after perihelion. The FAPOL mode does not employ a strip mask, causing the ordinary and extraordinary beams to overlap on the detector. The photocentres of the two beams in the resultant image are separated by 15'', meaning this setup is best suited for point-like targets or extended targets that do not surpass this separation. While this configuration proved suitable during the pre-perihelion campaign, when the coma remained sufficiently compact, the larger apparent size of the post-perihelion coma at smaller geocentric distances would have resulted in significant overlap between the two beams. The WeDoWo configuration avoids this problem by spatially separating the four polarised images into vignetted strips approximately 10'' in width. The trade-off is a reduced field of view, but with complete separation of the polarised images, making the configuration better suited to extended cometary comae.

Since WeDoWo occupies the filter wheel position of the ALFOSC R$_{\rm Bes}$~650\_130 filter used during the pre-perihelion campaign, the post-perihelion ALFOSC observations were instead obtained using the $r'_{\rm SDSS}$~618\_148 filter. The resulting difference in effective wavelength is small compared to the smooth wavelength dependence of 3I's polarisation and is therefore negligible compared to the measurement uncertainties. Observations with this configuration were repeated $1-3$ times per epoch. The final values and corresponding errors reported in Table~\ref{tablePQ} are the mean and standard deviation of the mean. A high-polarisation and zero-polarisation standard star were observed in addition to the comet to measure and account for instrumental effects, and pre-processing consistent of standard bias and flat-field correction. 

\section{Gas Contamination}
To estimate the gas contribution in our polarimetric data, we followed the procedure of \citet{ishiguro2016}, which assumes that the contribution from gas emission is more significant in a given band (in this case, B- and V-bands), but negligible in the R and I bands. Under this assumption, using the B-band as an example, an approximate gas image can be constructed as:
\begin{equation}
I_{\rm{gas}} = I_{\rm{B}} - k*I_{\rm{I}}
\end{equation}
where $I_{\rm{X}}$ is the deep-image intensity map for filter X and the scaling parameter $k$ is tuned such that the residual signals cancel in the differential image. The corresponding dust image may then be estimated as:
\begin{equation}
I_{\rm{dust}} = I_{\rm{R}} - I_{\rm{gas}}
\end{equation}

In principle, $k$ is chosen such that the dust continuum is removed from the gas image. Following \citet{ishiguro2016}, this may be achieved by measuring $k \sim I_{\rm{B}}/I_{\rm{I}}$ in regions expected to be dominated by dust, such as the dust tail or a large annuli surrounding the photocentre. The value of $k$ is refined iteratively until the residual gas image exhibits an approximately spherical coma with minimal tail-like residuals, while the dust image retains the asymmetric dust-tail morphology. This is because, in theory, dust particles may be accelerated by solar radiation pressure, whereas neutral gas molecules are less sensitive to radiation pressure and expand almost spherically. 

Once a suitable value of $k$ has been determined, the fractional contribution of gaseous emission and dust to the total measured flux can be estimated using 
\begin{align}
f^{\rm{g}}_{\rm{B}} &= \frac{\rm{F_B} - k*\rm{F_I}}{\rm{F_B}}    \\
f^{\rm{d}}_{\rm{B}} &= 1 - f^{\rm{g}}_{\rm{B}}
\end{align}
where $\rm{F_B}$ and $\rm{F_I}$ are the measured fluxes in the B and I filters, respectively, within the same aperture size used for the polarimetric measurements, i.e. 2000-6000~km. Although this approach is necessarily approximate, it provides an order-of-magnitude estimate of the relative contributions of dust-scattered continuum and gaseous emission to the measured flux within a given aperture. 

\begin{figure}
    \centering
    \includegraphics[width=0.5\linewidth]{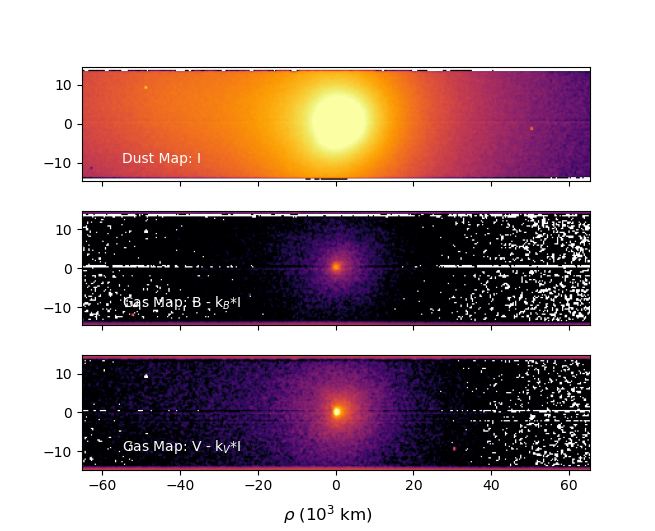}
    \caption{Approximate dust (\textit{top}) and gas (\textit{middle} and \textit{lower}) maps of 2025 December 26, obtained using image-subtraction technique described in text.}
    \label{Figure:GasContamination}
\end{figure}

We applied this procedure to each of our multi-band observations, measuring $k$ using various apertures positioned along the dust tail. Consequently, we found a range of plausible $k$ values that effectively removed the dust tail. Figure~\ref{Figure:GasContamination} shows an example of the differential images for the 2025 December 26 epoch data. 

Here, we report the mean values of the $f^{\rm{g}}_{\rm{B}}$ and $f^{\rm{g}}_{\rm{V}}$ gas-fraction estimates. For the 2025 December 18 epoch, we calculate $f^{\rm{g}}_{\rm{B}} = 5\%$ and $f^{\rm{g}}_{\rm{V}} = 9\%$. By the next multi-band epoch on December 26, we calculate slightly lower levels of $f^{\rm{g}}_{\rm{B}} = 3\%$ and $f^{\rm{g}}_{\rm{B}} = 6\%$. For the following multi-band epochs, the method did not produce robust constraints on the gas fraction, most likely because the gas emission had become too weak to distinguish reliably from the dust continuum. 

\begin{table*}
\caption{\label{tableLog} 3I/ATLAS observing log.}
\centering
\begin{tabular}{cccccccccccc}
\hline\hline
Date & UT & Instr. & $\alpha$ & $r_H$ & $\Delta$ & $\Phi$ & PA & Waveband & Exp & $n_Q$ & $n_U$ \\
     &    &        & (deg) & (au) & (au) & (deg) & (deg) &  & (s) &  &  \\
\hline\hline

2025-Nov-28 & 06:25 & ALFOSC & 30.4 & 1.7 & 1.9 & 293.8 & --- & r' & 150s & 1 & 1 \\

2025-Dec-17 & 08:03--08:17 & FORS2 & 25.5 & 2.2 & 1.8 & 292.6 & -23.6 & (V,R,F834) & 8x(1,1,4)s & 4 & 4 \\

2025-Dec-18 & 08:12--08:28 & FORS2 & 24.9 & 2.3 & 1.8 & 292.5 & -23.6 & (F485,V,R,F834) & 8x(5,1,1,4)s & 4 & 4 \\

2025-Dec-19 & 02:22 & ALFOSC & 24.4 & 2.3 & 1.8 & 292.4 & --- & r' & 3x150s & 3 & 3 \\

2025-Dec-26 & 01:52 & ALFOSC & 19.5 & 2.5 & 1.8 & 291.6 & --- & r' & 3x300s & 3 & 3 \\

2025-Dec-26 & 07:40--07:55 & FORS2 & 19.3 & 2.5 & 1.8 & 291.5 & -21.5 & (F485,V,R,F834) & 4x(20,8,6,25)s & 4 & 0 \\

2025-Dec-27 & 23:14 & FeReRo2 & 18.1 & 2.5 & 1.8 & 291.3 & --- & Rx & 8x3x300s & 6 & 6 \\

2025-Dec-28 & 22:49 & FeReRo2 & 17.3 & 2.6 & 1.8 & 291.2 & --- & Rx & 8x3x100s & 6 & 6 \\

2025-Dec-29 & 08:25 & FORS2 & 17.0 & 2.6 & 1.8 & 291.2 & -21.2 & R & 4x6s & 4 & 0 \\

2026-Jan-04 & 04:23--04:39 & FORS2 & 12.4 & 2.8 & 1.9 & 290.6 & -20.7 & (F485,V,R,F834) & 4x(30,12,8,30)s & 4 & 0 \\

2026-Jan-10 & 07:25--07:48 & FORS2 & 7.8 & 3.0 & 2.0 & 291.0 & -20.9 & (F485,V,R,F834) & 4x(80,20,15,65)s & 4 & 0 \\

2026-Jan-14 & 23:32 & ALFOSC & 4.6 & 3.1 & 2.1 & 293.4 & --- & r' & 3x350s & 3 & 3 \\

2026-Jan-15 & 23:38 & FeReRo2 & 4.0 & 3.1 & 2.2 & 294.5 & --- & Rx & 8x3x300s & 6 & 6 \\

2026-Jan-17 & 22:53 & FeReRo2 & 4.0 & 3.2 & 2.2 & 298.7 & --- & Rx & 8x3x300s & 6 & 6 \\

2026-Jan-18 & 05:19--05:42 & FORS2 & 2.6 & 3.2 & 2.2 & 299.6 & -30.0 & (F485,V,R,F834) & 4x(100,25,20,85)s & 4 & 0 \\

2026-Jan-20 & 02:59--03:28 & FORS2 & 1.5 & 3.3 & 2.3 & 310.7 & -41.0 & (F485,V,R,F834) & 4x(140,35,25,130)s & 4 & 0 \\

2026-Jan-20 & 18:33 & FeReRo2 & 1.2 & 3.3 & 2.3 & 320.1 & --- & Rx & 4x2x300s & 2 & 2 \\

2026-Jan-21 & 02:42--03:23 & FORS2 & 1.1 & 3.3 & 2.3 & 324.8 & 0.0 & (B,V,R,I) & 8x(90,35,25,35)s & 4 & 4 \\

2026-Feb-07 & 01:18--01:54 & FORS2 & 6.4 & 3.9 & 2.9 & 93.8 & -3.8 & (B,V,R,I) & 4x(210,85,60,85)s & 4 & 0 \\

\hline \hline
\end{tabular}
\vspace{0.5cm}
\noindent \parbox{\linewidth}{
\textbf{Note.} The UT column represents the midpoint of each observing sequence; for multi-band cases, the midpoints of the first and last filter observations are given. $\alpha$ is the phase angle; $r_H$ the heliocentric distance; $\Delta$ the geocentric distance, and $\Phi$ the position angle of the scattering plane. PA denotes the instrument rotation angle (relevant only for FORS2). Exp gives the exposure times for each filter, ordered as listed in the Waveband column. $n_Q$ and $n_U$ indicate the number of frames used to compute Stokes Q and U, respectively. 
}
\end{table*}

\section{Database of Cometary Polarimetry}
We compiled multi-band polarimetric data from the literature to compare the polarimetric colour of 3I with that of other comets. We first searched the Database of Cometary Polarimetry (DBCP)\citep{kiselev2017} for contemporaneous multi-band observations, i.e., measurements obtained by the same observing group on the same night in wavebands comparable to our BVRI filters. Since our aim is to investigate not only the polarimetric colour, but also its rate of change with phase angle, we restricted the comparison to objects with datasets spanning a sufficiently broad range of phase angles, with coverage extending to phase angles somewhat comparable to those sampled by our observations. 

Based on these criteria, we selected observations of 1P/Halley \citep{kikuchi1987,visvanathan1991,sen1991}, 67P/Churyumov-Gerasimenko \citep{hadamcik2010,gray2024MNRAS,rosenbush2026}, 109P/Swift-Tuttle \citep{kikuchi1987}, C/Hale-Bopp \citep{kikuchi1987,kiselev1997,ganesh1998,manset2000}, and 2I/Borisov \citep{bagnulo2021Natur}. For each dataset, we calculated the polarimetric colour and its variation with phase angle, $\Delta \rm{PC}(\lambda_1,\lambda_2)/\alpha$, following the same procedure described in Section \ref{Sect_PolCol}.  

\end{document}